\input{aipcheck}

\documentclass[
    ,final            % use final for the camera ready runs
  ]
  {aipproc}

\layoutstyle{6x9}

\begin{document}

\title{Perturbative and Nonperturbative Aspects of Azimuthal 
Asymmetries in Polarized $ep$~\footnote{Presented at the 15th International 
Spin Physics Symposium (Spin 2002), Brookhaven National Laboratory, 
September 9-14, 2002.}}

\author{K. A. Oganessyan}{
  address={LNF-INFN, I-00040, Enrico Fermi 40, Frascati, Italy}
  ,address={DESY, Notkestrasse 85, 22603 Hamburg, Germany}
}

\begin{abstract}
We discuss the possibilities of testing of perturbative quantum 
chromodynamics through azimuthal asymmetries in polarized $ep$ 
in the context of future collider and fixed target facilities, 
such as Electron Ion Collider and TESLA-N.    
\end{abstract}

\maketitle

It is widely recognized that the study of the distributions in the 
azimuthal angle $\phi$ of the detected hadron in hard scattering 
processes provide interesting variables to study in both  
non-perturbative and perturbative regimes. They are of great 
interest since they test perturbative quantum chromodynamics 
(QCD) predictions for the short-distance part of strong 
interactions and yield an important information on the 
long-distance internal structure of hadrons which computed 
in QCD by non-perturbative methods.  

We discuss here the perturbative aspects of spin-independent, 
single- and double-spin azimuthal symmetries in leptoproduction 
processes. The non-perturbative aspects of some of them have been    
discussed in these proceedings (see Refs.~[1-5]). 

\subsection{
Spin-independent $cos\phi$, $cos2\phi$ asymmetries 
}

Different mechanisms to generate spin-independent azimuthal asymmetries 
in semi-inclusive processes have been discussed in the literature. 
Georgi and Politzer \cite{PG} found a negative contribution to 
$\langle \cos\phi \rangle$ in the first order in $\alpha_S$ perturbative 
theory and proposed the measurement of this quantity as a clean test 
of QCD. However, as Cahn~\cite{CAHN} showed, there is a contribution 
to $\langle \cos\phi \rangle $ from the lowest-order processes due to 
this intrinsic transverse momentum. The measurements indicate  
large negative $\cos\phi$ asymmetry~\cite{EMC,E665} and the simple 
QCD-improved parton model rather well describes the data, where 
essential contribution comes from non-perturbative intrinsic 
transverse momentum effects.   

Recently, the expectation values of $\cos\phi$ and $\cos2\phi$ in the 
deep inelastic electroproduction of single particles in the high $Q^2$ 
region have been observed by ZEUS detector at HERA~\cite{ZEUS} for the 
first time. For hadrons produced at large transverse momenta the 
results are in agreement with QCD predictions~\cite{TG0,OG,NSY}. Since 
the non-perturbative contributions to $\langle \cos2\phi\rangle$ 
are of $1/Q^2$, its precise 
measurement may provide clear evidence for a pQCD and an alternative 
possibility to measure $F_L(x,Q^2)$~\cite{TG}. 

\subsection{
Double-spin $cos\phi$, $cos2\phi$ asymmetries
} 

The $\cos\phi$ and $\cos2\phi$ asymmetries arise also in semi-inclusive 
DIS of longitudinally polarized electrons off longitudinally polarized 
protons. Recently, the non-perturbative phenomena of these asymmetries 
has been considered~\cite{OMD} and a sizable negative $\cos\phi$ 
asymmetry for $\pi^{+}$ electroproduction was predicted. 
Perturbative contributions proportional to $\alpha_s(Q^2) \ldots g_1 \ldots$ 
will likely appear at the same point where the twist-3 function 
$(M/Q)\ldots g_L^\perp$ appears~\cite{OMD,TM,BMO}, like contributions 
proportional to $\alpha_s(Q^2) \ldots f_1 \ldots$ appear at the same 
point where the function $(M/Q)\ldots f_1^\perp$ appears~\cite{PG,OMD,TM}. 
Then the $\cos\phi$ dependence of the double longitudinal spin asymmetry, 
\begin{equation}
A_{LL} = \frac{d\sigma^{++}+d\sigma^{--}-d\sigma^{+-}-d\sigma^{-+}}
{d\sigma^{++}+d\sigma^{--}+d\sigma^{+-}+d\sigma^{-+} },
\label{AS}
\end{equation}
for charged pion electroproduction have been examined~\cite{OAAD}. 
The subscript $LL$ denotes the longitudinal polarization of the beam 
and target, and the superscripts $++,-- (+-,-+)$ denote 
the helicity states of the beam and target respectively, corresponding 
to antiparallel (parallel) polarization.   

In general, due to parity conservation in the electromagnetic and strong 
interactions, Eq.(\ref{AS}) can be written in terms of the spin-independent 
($\sigma$) and double-spin ($\Delta \sigma$) cross sections of 
semi-inclusive DIS 
\begin{equation}
\label{AS1}
A_{LL} =\frac{2\,\sum_{m=1} \Delta \sigma^{m}_{LL} 
\cos {([m-1] \cdot \phi)}}{4\, \sum_{m=1} \sigma^{m}_{UU} 
\cos {([m-1] \cdot \phi)}}, 
\end{equation}
Up to sub-leading order $1/Q$ the Eq.(\ref{AS1}) can be rewritten as  
\begin{equation}
A_{LL} = \frac{\Delta \sigma^{1}_{LL}/2 \sigma^{1}_{UU} + 
{\langle \cos\phi \rangle}_{LL} 
\cdot \cos\phi}{1 + 2\,{\langle \cos\phi \rangle}_{UU} \cdot \cos\phi } ,  
\label{AP}
\end{equation}
where ${\langle \cos\phi \rangle}_{UU}$ and ${\langle \cos\phi \rangle}_{LL}$ 
are unpolarized and double polarized $\cos\phi$ moments, respectively. 
The $A_{LL}$ depends on and allows the simultaneous determination of 
the spin-independent and double-spin $\cos\phi$-moments of the cross 
section. A $\cos2\phi$ asymmetry only appears at order $1/Q^2$, unless 
one allows for T-odd structure functions~\cite{DB,LG}. 

\subsection{
Beam/target single-spin azimuthal asymmetries 
}
The interest in the single beam asymmetry in the pion electroproduction
in semi-inclusive deep inelastic scattering of longitudinally polarized
electrons off unpolarized nucleon resides in the fact that they probe the
antisymmetric part of the hadron tensor, which entirely due and particularly
sensitive to final state interactions (FSI). In longitudinally polarized
electron scattering it shows up as a $<\sin\phi>$ asymmetry for the produced 
hadron and expressed as 
\begin{equation}
\langle \sin\phi \rangle\,=\, \pm\, \langle \frac{\vec{s} \times \vec{k}' 
\cdot \vec{P}_{h\perp}}{\vert \vec{s} \times \vec{k}' \vert \vert 
\vec{P}_{h\perp} \vert} \rangle,
\label{A1}
\end{equation}
where $\vec{s}$ denoted the spin vector of the electron (the upper (lower)
sign for right (left) handed electrons), $\vec{k}$ ($\vec{k}'$) and 
$\vec{P}_{h\perp}$ are three vectors of incoming (outgoing) electron and the 
produced hadrons transverse momentum about virtual photon direction. 
This asymmetry is related to the left-right asymmetry in the hadron 
momentum distribution with respect to the electron scattering plane,
\begin{equation}
A = \frac{\int_0^\pi d\phi d\sigma
- \int_\pi^{2\pi} d\phi d\sigma}{\int_0^\pi d\phi d\sigma
+ \int_\pi^{2\pi} d\phi d\sigma},
\label{A2}
\end{equation}
which is $4/\pi$ times $<\sin\phi>$. 

Among various proposed tests to measure the gluon self-coupling, 
the tests based on observation of time-reversal-odd (T-odd) 
processes, which are highly sensitive to FSI, are particularly promising. 
In the one-loop order, which is the leading order for T-odd 
quantities in pQCD~\cite{RPL}, the left-right asymmetry is quite 
sensitive to the gluon self-coupling~\cite{HAG}, and even the qualitative 
observation of such effects is sufficient to establish its existence. 
Furthermore, they are fairly insensitive to the poorly known gluon 
distribution and fragmentation functions~\cite{HAG}. The magnitude of 
this asymmetry for kinematical conditions (assuming $\sqrt{s} = 300$ GeV) 
at HERA collider have been evaluated using parton model expressions at 
leading order~\cite{AG}. Although the asymmetry appears small at first 
sight, their measurement may still 
be possible. An experimental determination of it is therefore a 
challenging task. In particular by including only hadrons above a 
minimal transverse momentum in the measurement, the 
asymmetries can be suppressed to $\cal{O}$ $(\alpha_S^{(1)})$. It is 
worthy to not that in the electromagnetic scattering the difference in 
magnitude between the opposite helicity measurements and the asymmetry 
$<\sin2\phi>$ is proportional the parity-violating effects.           
 
Within the last few years, after appearance of experimental results 
from  HERMES~\cite{HERM,HERM1} and SMC~\cite{SMC} collaborations on 
single target-spin azimuthal asymmetries in semi-inclusive pion 
electroproduction the subject obtained much more attention. It is 
considered as one of the ways to access and measure third, completely 
unknown distribution function, so-called, transversity. This twist-2 
function is equally important for the description of quarks in nucleons 
as the more familiar function $g_1(x)$; their information is 
complementary (for a review of transversity see Refs.~\cite{BDR,RATC}). 

In the perturbative regime the single target-spin asymmetries also 
should be different from zero in the one-loop order. They are remaining 
out of attention, while they may give a valuable information about the 
behavior of QCD interactions under discrete symmetry operations and 
may test our understanding of FSI at the perturbative level.  

\vspace*{0.7cm}

The future collider and fixed target facilities, such as Electron 
Ion Collider (EIC)~\cite{EIC} and TESLA-N~\cite{TN} will be very 
important for deeper study of considered azimuthal asymmetries. 
In particular, the wide and tunable range of collision energies 
at EIC will allow to study the nonperturbative and perturbative 
effects of azimuthal asymmetries consistently. Polarization of 
electron and proton spins will allow to measure all discussed 
asymmetries as well as their different combinations simultaneously. 
Some of the asymmetries are expected to be small, thus the facilities 
with high luminosity, such as EIC, TESLA-N are required. 

From the theoretical side, a realistic estimates of these azimuthal 
asymmetries for the future collider and fixed target facility 
kinematics are needed. 
 
\vspace*{1cm}

I am grateful to Abhay Deshpande and Werner Vogelsang for many valuable 
discussions and the Physics Department at Brookhaven National Laboratory 
for hospitality.

\end{document}